\def   \ni {\noindent}

\def   \ssk {\vskip  5truept}

\def   \bsk {\vskip 15truept}
 
\def   \newpage {\vfill\eject}
\def   \newline {\hfil\break}

\documentstyle[epsfig]{article}
\begin{document}

\hsize 5truein
\vsize 8truein
\font\abstract=cmr8
\font\keywords=cmr8
\font\caption=cmr8
\font\references=cmr8
\font\text=cmr10
\font\affiliation=cmssi10
\font\author=cmss10
\font\mc=cmss8
\font\title=cmssbx10 scaled\magstep2
\font\alcit=cmti7 scaled\magstephalf
\font\alcin=cmr6 
\font\ita=cmti8
\font\mma=cmr8
\def\ref{\par\noindent\hangindent 15pt}
\null
%\vskip 3.0truecm
%\baselineskip = 12pt

% ------ beginning of font "title" ------

\title{\ni GAMMA-RAY EMISSION OF CLASSICAL NOVAE\\ 
           AND ITS DETECTABILITY BY INTEGRAL}

% beginning of font "author and affiliation"
\bsk \bsk
\author{\ni M.~Hernanz $^{1}$, J.~G\'omez-Gomar $^{1}$, J. Jos\'e $^{1,2}$,
         A. Coc $^{3}$, J. Isern $^{1}$}
\bsk
\affiliation{1) Institut d'Estudis Espacials de Catalunya IEEC/CSIC/UPC, 
             Edifici Nexus, C/Gran Capit\`a 2-4, 08034 Barcelona, SPAIN,
             2) Departament de
             F\'{\i}sica i Enginyeria Nuclear, UPC, Avda. V\'{\i}ctor
             Balaguer s/n, 08800 Vilanova i la Geltr\'u, Barcelona, Spain,
             3) Centre de Spectrom\'etrie Nucl\'eaire et de Spectrom\'etrie 
	     de Masse, CSNSM, IN2P3-CNRS, B\^at. 104, 91405 Orsay, France
}                                                
\bsk
\baselineskip = 12pt

% beginning of font "abstract and keywords"
\abstract{ABSTRACT \ni
A lot of information concerning the mechanism of nova explosions will be
extracted from the possible future observations with INTEGRAL. In order
to be prepared for this task, we are performing a detailed modelization
of the $\gamma$-ray emission of classical novae, for a wide range of possible
initial conditions. New models of classical novae have been computed with a
continuously updated hydrodynamical code, including a complete network of
nuclear reactions. 
Spectra at different epochs after the explosion and light curves for the
different lines (511, 478 and 1275 keV) and the continuum are presented, as 
well as detectability distances with INTEGRAL's spectrometer SPI. 
}                                                    
\bsk
\baselineskip = 12pt

\bsk
\baselineskip = 12pt

% beginning of font "text"

\text{\ni 1. INTRODUCTION
\ssk
\ni     

Accretion of hydrogen-rich matter by a white dwarf, in a close binary system, 
is at the origin of nova outbursts. The explosion is triggered by the 
thermonuclear runaway (TNR) that ensues from hydrogen ignition in degenerate 
conditions at the base of the accreted envelope. 
During nova explosions, some radioactive nuclei 
are synthesized which should produce emissions 
of $\gamma$-rays. Both line and continuum emission are expected. The 
annihilation 
of the positrons coming from the $\beta^+$--decay of $^{13}$N and $^{18}$F 
($\tau$=862s and 158 min, respectively) produces emission at 511 keV and below, 
whereas the decays of $^7$Be (through an e$^-$--capture) and $^{22}$Na
(through a $\beta^+$--decay) emit 478 and 1275 keV photons, with lifetimes 
of 77 days and 3.75 years, respectively. 

A detailed analysis of the $\gamma$-ray emission of classical novae requires 
the 
coupling of a hydrodynamical code, which gives the detailed nucleosynthesis of 
the radioactive nuclei, as well as the temporal evolution of the properties of 
the expanding envelope, to a Monte-Carlo one, able to handle $\gamma$-ray 
production and transfer. The potential importance of $\gamma$-ray emission of 
classical novae had been already pointed out in previous works (Clayton \& 
Hoyle 1974, Clayton, 1982, Leising \& Clayton 1987).

\bsk
\ni 2. GAMMA-RAY EMISSION OF INDIVIDUAL CO AND ONe NOVAE
\ssk
\ni 
\newpage

From the observational point of view, two main types of novae can be 
distinguished: {\it standard} novae and {\it neon} novae. All of them show 
enhancement in CNO elements, but {\it neon} novae are also particularly 
enhanced in
neon. It is largely accepted that {\it standard} novae are occurring on CO 
white 
dwarfs, whereas {\it neon} novae are exploding ONe white dwarfs; in both
cases, some initial enrichment of the accreted material through mixing with 
material from the core (either via diffusion, dredge-up or other mechanisms)
is needed in order to explain the observed abundances and even the explosive 
process itself. Thus, the initial composition of the accreted material plays 
an important role in the outburst properties (Jos\'e \& Hernanz 1998). 
Concerning the $\gamma$-ray 
emission, we have demonstrated in previous works that there is a clear 
distinction between CO 
and ONe novae. CO novae show emission at 478 keV, related to $^7$Be--decay, 
lasting for 2 months, 
whereas ONe novae show emission at 1275 keV, related to $^{22}$Na decay, 
lasting for some years (see Hernanz et al., 1996 and G\'omez-Gomar et al. 
1998 for details). This is 
a consequence of the different nucleosynthetic yields of CO and ONe novae 
(Jos\'e \& Hernanz, 1998, Kovetz \& Prialnik, 1997, Starrfield et al. 1998).  

An example of the $\gamma$-ray emission of CO novae is shown in Figure 1, where 
the light curves of the 478 keV line for 0.8 and 1.15 M$_\odot$ novae are 
shown. The width of the lines (3 and 7 keV) has been taken into account for 
the determination of the maximum detectability distance of the line by the 
future spectrometer SPI onboard INTEGRAL; this distance is around 0.5 kpc. 

Concerning 1275 keV emission, new results since our previous works
are shown in Table 1 and Figure 2. These correspond to a reanalysis of the 
rates of the nuclear reactions involved in the synthesis of $^{22}$Na and 
$^{26}$Al in ONe novae (Jos\'e, Coc \& Hernanz, 1998), 
which are the main novae producers of these elements, mainly because 
of their initial enrichments in $^{20}$Ne, $^{23}$Na and $^{24,25}$Mg. 
The new recommended rates of some key reactions of 
the NeNa-MgAl cycles and the upper and lower limits of them, translate into 
a range of ejected masses of $^{22}$Na, which 
in all cases are larger than the old ones (see Table 1). The much lower 
$^{22}$Na yield 
of a CO nova is also shown for comparison. 
The light curves of the 1275 keV line are correspondingly shifted, as is 
shown in Figure 2 for an ONe nova of 1.25 M$_\odot$. The new 1275 keV fluxes
are larger by, at most, a factor of $\sim$10. Taking into account that the 
maximum detectability distances for ONe novae (for INTEGRAL's spectrometer
SPI) were around 0.5 kpc for the old 
models (see G\'omez-Gomar et al. 1998), we deduce that in the most optimistic 
case detectability distances can be as large as 2 kpc, without taking into 
account the effect of larger ejected masses, which is a long standing problem 
of all theoretical models (i.e., the unability to reproduce some large 
ejected masses observed; see Starrfield et al. 1998, for a recent discussion
on this topic). The width of the lines is $\sim$20 keV in this case. Up to 
now, the COMPTEL instrument onboard the Compton GRO has 
established upper limits to the 1275 keV emission from some recent novae, which 
are compatible with our computed fluxes (Iyudin et al. 1995). For an 
analysis of the cumulative 1275 keV emission from novae in the Galaxy, see 
Jean et al. 1998 (and these proceedings).

Another important emission of classical novae in the $\gamma$-ray domain is 
the 511 keV line and the continuum below it. This emission, which is produced 
both in CO and in ONe novae, is by far the most 
intense one, but it has a very short duration. The reason 
is that it is related mainly to the disintegration of $^{13}$N and $^{18}$F, 
which
decay emitting positrons on very short timescales (see above). Thus, only for 
a very early detection (even before the visual maximum for the 
majority of the cases) would this emission be detected up to distances of 
$\sim$10 kpc (Hernanz et al. 1996, G\'omez-Gomar et al. 1998). In this sense, 
the use of the shield of the INTEGRAL spectrometer as a nova 
detector in $\gamma$-rays would be of great importance (see Jean et al.,
these proceedings). However, there are very recent results concerning nuclear 
reaction rates related to $^{18}$F destruction by (p,$\gamma$) and 
(p,$\alpha$) reactions (Utku et al. 1998) which will significantly affect 
$^{18}$F synthesis in novae. Our still preliminary computations indicate that 
$^{18}$F production will be lowered by a factor $\sim$10, leading to smaller 
511 keV line fluxes and smaller detectability distances (the 
complete calculation is in progress). 

%---------- table 1
\begin{table}
\centering
\caption{TABLE 1: $^{22}$Na ejected masses (in M$_\odot$).} \\ 
\begin{tabular}{@{}lccccc} \hline \hline
Type  & M$_{WD}$ (M$_\odot$)  & M$_{\rm ejec}$(old) & M$_{\rm ejec}$(high) & 
        M$_{\rm ejec}$(low) \\ \hline
ONe   & 1.15                  & 1.0 10$^{-9}$       & 1.2 10$^{-8}$       & 
        3.1 10$^{-9}$          \\
ONe   & 1.25                  & 1.3 10$^{-9}$       & 9.0 10$^{-9}$       & 
        2.9 10$^{-9}$          \\
ONe   & 1.35                  & 2.6 10$^{-9}$       & 6.2 10$^{-9}$       & 
        3.4 10$^{-9}$          \\ 
CO    & 1.15                  & 3.8 10$^{-12}$      &  --                 & 
        --                     \\ \hline \hline
\end{tabular}
\end{table}
%--------------------------  figure 1
\begin{figure}
\centerline{\psfig{file=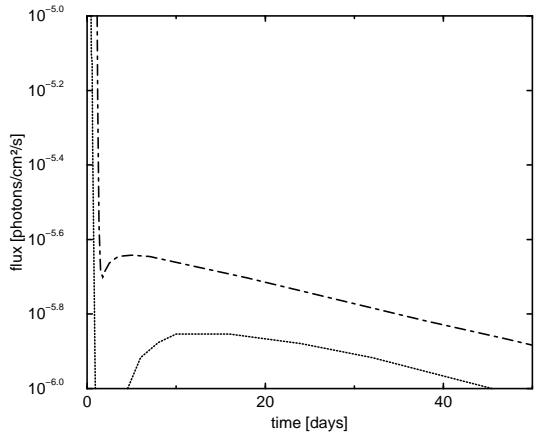, width=7cm}}
\centering
\caption{FIGURE 1. Light curves for the 478 kev line (D=1kpc) for CO novae of 
0.8M$_\odot$ (dotted line) and 1.15M$_\odot$ (dash-dotted line).}
\end{figure}
%--------------------------  figure 2
\begin{figure}
\centerline{\psfig{file=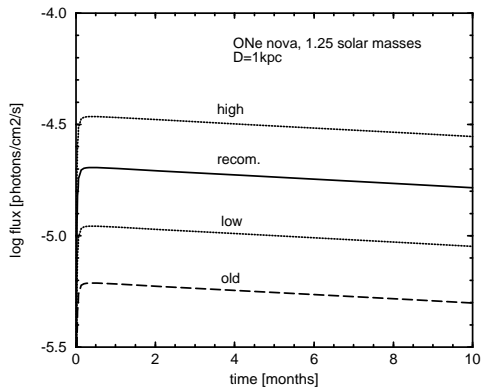, width=7cm}}
\centering
\caption{FIGURE 2. Light curves for the 1275keV line} 
\end{figure}

\bsk
\ni 3. CONCLUSIONS 
\ssk
\ni 

Novae are potential $\gamma$-ray emitters, mainly through continuum emission, 
between $\sim$30 and 511 keV, and line emission at 511 keV (all types of
novae). This emission could be detected with SPI, provided that the novae 
are caught early enough (even before the maximum in visual light). 
Line emission at 478 keV (CO novae) and 1275 keV (ONe novae) is less intense 
than continuum and 511 keV line emission, but has much longer duration,
specially the 1275 keV one. Its detection by INTEGRAL would provide an 
important confirmation of the TNR model of classical novae and insights on
the conditions in the expanding ejecta.

}

\bsk
\baselineskip = 12pt
{\abstract \ni ACKNOWLEDGMENTS

This research has been partially supported by the DGICYT
(PB94-0827-C02-02), CICYT (ESP95-091), CIRIT, AIHF908 and PIC 319
}

\bsk
\baselineskip = 12pt

% beginning of font "references"

{\references \ni REFERENCES
\ssk

\ref Clayton, D.D. 1981, ApJ, 244, L97
\ref Clayton, D.D., Hoyle, F. 1974, ApJ, 187, L101
\ref G\'omez-Gomar, J., Hernanz, M., Jos\'e, J., Isern, J. 1988, MNRAS, 296,
913
\ref Hernanz, M., G\'omez-Gomar, J., Jos\'e, J., Isern, J. 1997, in 
Proceedings of the 2$^{nd}$ INTEGRAL Workshop The Transparent Universe, 
ESA SP-382, p.47
\ref Iyudin et al. 1995, A\&A, 300, 422
\ref Jean, P., Hernanz, M., G\'omez-Gomar, J., Jos\'e, J. 1998, in 
preparation
\ref Jos\'e, J., Hernanz, M. 1998, ApJ, 494, 680
\ref Jos\'e, J., Coc, A., Hernanz, M. 1998, ApJ, submitted
\ref Kovetz, A., Prialnik, D. 1997, ApJ, 477, 356
\ref Leising, M.D., Clayton, D.D. 1987, ApJ, 323, 159
\ref Starrfield, S., Truran, J.W., Wiescher, M.C., Sparks, W.M. 1998, 
MNRAS, 296, 502
\ref Utku, S. et al. 1998, Phys. Rev. C, 57, 2731
}                      

\end{document}